\documentclass[aps,twocolumn]{revtex4}
\usepackage{amsmath}
\usepackage{graphicx}
\usepackage{floatflt,epsfig}
\usepackage{lscape}
\def\ii{\'{\i}}
\def\beq{\begin{equation}}
\def\eeq{\end{equation}}
\def\beqa{\begin{eqnarray}}
\def\eeqa{\end{eqnarray}}
\def\ban{\begin{eqnarray*}}
\def\ean{\end{eqnarray*}}
\def\bi{\begin{itemize}}
\def\ei{\end{itemize}}

\def\gsim{~\rlap{$>$}{\lower 1.0ex\hbox{$\sim$}}}
\begin{document}

\title{Collective modes in relativistic npe matter at finite temperature}

\author{ L. Brito, C. Provid\^encia, A. M. Santos}
\affiliation{Centro de F\ii sica Te\'orica - Departamento de F\ii sica\\
Universidade de Coimbra - P-3004 - 516 - Coimbra - Portugal}
\author{S. S. Avancini, D. P. Menezes}
\affiliation{Depto de F\'{\i}sica - CFM - Universidade Federal de Santa
Catarina\\
Florian\'opolis - SC - CP. 476 - CEP 88.040 - 900 - Brazil}
\author{Ph. Chomaz}
\affiliation{GANIL (DSM-CEA/IN2P3-CNRS), B.P. 5027, F-14076 Caen C\'edex 5,
France}

\begin{abstract}
Isospin and density waves in neutral neutron-proton-electron (npe) matter are 
studied
within a relativistic mean-field hadron model at finite temperature
with the inclusion of the electromagnetic field. The dispersion relation
  is calculated and the collective modes are obtained. The unstable modes are
  discussed and the spinodals, which separate the stable from the unstable 
regions, are shown for different values of the momentum transfer at various
temperatures. The critical temperatures are compared with the ones obtained in
a system without electrons. The largest critical temperature, 12.39 MeV,
occurs for a proton fraction $y_p=0.47$. For  $y_p=0.3$ we get $T_{cr}$=5 MeV
and for  $y_p>0.495$ $T_{cr}\lesssim 8$ MeV.
 It is shown that at  finite temperature 
 the distillation effect in asymmetric matter  is not so efficient and
 that electron effects are particularly important for small momentum transfers.

PACS number(s): {21.60.-n,21.60.Ev,21.65.+f,24.10.Jv,71.10.Ay}
\end{abstract}

\maketitle

\section{Introduction}

In order to describe compact-star matter at low densities, besides
protons and neutrons,  the electrons (and possibly neutrinos if they
are trapped)  must be introduced. 
Electrons neutralize the proton charge and thus suppress the diverging Coulomb 
contribution to the energy. This composite matter is expected 
to undergo phase transitions associated to the nuclear 
liquid-gas phase transition but strongly modified by the Coulomb 
interaction and the presence of electrons.  
Phase transitions in asymmetric matter are related with the instability
  regions limited by the spinodal sections \cite{instabilities1,previous}.
These very same instabilities are responsible for nucleation processes
\cite{gotas} and the possible existence of a non-homogeneous phase.
At low nuclear matter densities, which are the densities of relevance in the 
present work, a competition between the long-range Coulomb repulsion and 
short-range nuclear attraction can lead to the formation of matter known as 
nuclear pasta \cite{pasta}. The competition of these two forces, normally 
acting on different scales,
is called frustration. This novel state of matter, the nuclear pasta, can 
appear in different structures and its properties have important consequences
in the crust of neutron stars and in the core-collapse of supernova 
\cite{horo2,wata04,maru,hpbp}.
Not only the equation of state of stellar matter has to be 
understood, but also the neutrino mean free path in the medium has to be
well described. In fact, neutrino
interactions are crucial in the dynamics of the core-collapse supernovae 
because they carry most of the energy away.
 It has been shown that the neutrino opacity is affected
by nucleon-nucleon interactions  due to coherent scattering off density
fluctuations \cite{sawyer}. Both single particle and collective
contribution have to be taken into account.  In \cite{rbp} it
was shown  that coherent neutrino scattering from non-uniform 
hadron-quark matter or hadron matter with and without kaon condensed phase 
would greatly reduce the neutrino mean-free path. A similar effect at low 
densities could allow enough energy transfer in order to revive the supernova 
shock.
Recent semiclassical simulations of the  linear response of nuclear 
non-homogenous matter at low densities, the so called nuclear pasta,  
to  neutrinos  in \cite{hpbp,horo2} have shown that coherence effects reduce 
the mean free path of neutrinos. In these simulations
electrons are not modeled explicitly, but their effect is included 
through a modified Coulomb interaction between the protons through  a 
screening length. Neutrinos will loose energy by 
exciting collective nuclear  modes or plasmon modes. However in
\cite{previous1} it has been shown that the behavior of the electrons 
depends on the wavelength of the perturbation.  In \cite{ksg} low energy 
nuclear collective excitations of Wigner-Seitz cells 
containing nuclear clusters immersed in a gas of neutrons have been obtained. 
In that paper, the electron motion was not included. Including the electron 
contribution will likely affect the results for large clusters.

 In  \cite{previous1} we have  investigated the influence of the
electromagnetic interaction and  the presence of electrons 
on  the unstable modes of nuclear matter  and compared the dynamical
instability region with the thermodynamical one.
 This investigation was performed in the framework of a
relativistic mean field hadronic model within the Vlasov
formalism \cite{npp91,npp93,previous}. 
In particular, in \cite{previous1}, it has been  studied 
the role of isospin and the modification of the distillation 
phenomenon due to the presence of the Coulomb field and electrons. These 
calculations  were performed at $T=0$ MeV.

In the present work the effect of 
dynamical instabilities including the Coulomb field are investigated at
finite $T$. Understanding the mixed phase of neutral matter 
at finite temperature is important to determine the behavior of neutrinos 
emitted in a supernova explosion.

\section{The Vlasov equation formalism}

We consider a system of baryons, with mass $M$
interacting with and through a isoscalar-scalar field $\phi$ with mass
$m_s$,  a isoscalar-vector field $V^{\mu}$ with mass
$m_v$ and an isovector-vector field  $\mathbf b^{\mu}$ with mass
$m_\rho$. We also include a system of electrons with mass $m_e$. Protons and
electrons
interact through the electromagnetic field $A^{\mu}$.
The lagrangian density reads:

\begin{equation}
\mathcal{L}=\sum_{i=p,n}\mathcal{L}_{i}+\mathcal{L}_e\mathcal{\,+L}_{{\sigma }}%
\mathcal{+L}_{{\omega }}\mathcal{+L}_{{\rho }}\mathcal{+L}_{A},
\label{lagdelta}
\end{equation}
where the nucleon Lagrangian reads
\begin{equation}
\mathcal{L}_{i}=\bar{\psi}_{i}\left[ \gamma _{\mu }iD^{\mu }-M^{*}\right]
\psi _{i}  \label{lagnucl},
\end{equation}
with 
\begin{eqnarray}
iD^{\mu } &=&i\partial ^{\mu }-g_{v}V^{\mu }-\frac{g_{\rho }}{2}{\vec{\tau}}%
\cdot \vec{b}^{\mu } - e \frac{1+\tau_3}{2}A^{\mu}, \label{Dmu} \\
M^{*} &=&M-g_{s}\phi,
\label{Mstar}
\end{eqnarray}
and the electron  Lagrangian is given by
\begin{equation}
\mathcal{L}_e=\bar \psi_e\left[\gamma_\mu\left(i\partial^{\mu} + e A^{\mu}\right)
-m_e\right]\psi_e.
\label{lage}
\end{equation}
The isoscalar part is associated with the scalar sigma ($\sigma $) field, $%
\phi $, and the vector omega ($\omega $) field, $V_{\mu }$, while the
isospin dependence is coming from the isovector vector rho ($\rho $) field, $b_{\mu
}^{i}$ (where $\mu $ is the 4 dimensional space-time indices and $i$ the 3D
isospin direction indices). The associated Lagrangians are 
\begin{eqnarray*}
\mathcal{L}_{{\sigma }} &=&+\frac{1}{2}\left( \partial _{\mu }\phi \partial %
^{\mu }\phi -m_{s}^{2}\phi ^{2}-\frac{1}{3}\kappa \phi ^{3}-\frac{1}{12}%
\lambda \phi ^{4}\right)  \\
\mathcal{L}_{{\omega }} &=&-\frac{1}{4}\Omega _{\mu \nu }\Omega ^{\mu \nu }+%
\frac{1}{2}m_{v}^{2}V_{\mu }V^{\mu }\\
%+\frac{1}{4!}\xi g_{v}^{4}(V_{\mu}V^{\mu })^{2} \\
\mathcal{L}_{{\rho }} &=&-\frac{1}{4}\vec{B}_{\mu \nu }\cdot \vec{B}^{\mu
\nu }+\frac{1}{2}m_{\rho }^{2}\vec{b}_{\mu }\cdot \vec{b}^{\mu }\\
\mathcal{L}_{A} &=&-\frac{1}{4}F _{\mu \nu }F^{\mu
  \nu }
\end{eqnarray*}
where $\Omega _{\mu \nu }=\partial _{\mu }V_{\nu }-\partial _{\nu }V_{\mu }$
, $\vec{B}_{\mu \nu }=\partial _{\mu }\vec{b}_{\nu }-\partial _{\nu }\vec{b}%
_{\mu }-g_{\rho }(\vec{b}_{\mu }\times \vec{b}_{\nu })$ and $F_{\mu \nu }=\partial _{\mu }A_{\nu }-\partial _{\nu }A_{\mu }$.
The model comprises the following parameters:
three coupling constants $g_s$, $g_v$ and $g_{\rho}$ of the mesons to
the nucleons, the nucleon mass $M$, the electron mass $m_e$, the masses of
the mesons $m_s$, $m_v$, $m_{\rho}$, the electromagnetic coupling constant
$e=\sqrt{4 \pi/137}$ and
the self-interacting coupling constants $\kappa$ and $\lambda$.
We have used the  set of constants identified as NL3 taken
from \cite{nl3}. For this case, the saturation density that we refer as
  $\rho_0$ is 0.148 fm$^{-3}$.
In the above lagrangian density $\vec{\tau}$ is
the isospin operator.

 Denoting by
$$f({\bf r},{\bf p},t)_\pm=\mbox{diag}(f_{p \pm},\,f_{n \pm},\,f_{e \pm})$$
the  distribution functions of particles (+) at position $\mathbf r$,
instant $t$ and momentum $\mathbf{p}$ and of antiparticles (-) at position $\mathbf r$,
instant $t$ and momentum $-\mathbf{p}$
and the corresponding one-body hamiltonian by
\begin{equation}
h_{\pm}= \mbox{diag}\left(h_{p\pm},h_{n\pm},h_{e\pm}\right)
\label{aga}
\end{equation}
where
$$h_{i\pm}=\pm \sqrt{({\bf p}-{\boldsymbol{\cal V}_i})^2 + {M^*}_i^2} +
{\cal V}_{0i}.$$
For protons and neutrons, $i=p,n$, we have
$${\cal V}_{0i}= g_v V_0  + \frac{g_\rho}{2}\, \tau_i b_0+ e\, A_0
\frac{1+\tau_{i}}{2}, $$
$${\boldsymbol{{\cal V}_{i}}}= g_v {\mathbf V} +
\frac{g_\rho}{2}\, \tau_i {\mathbf b}+ e\,  {\mathbf A}
\frac{1+\tau_{i}}{2},$$
$$M^*_p=M^*_n=M^*=M-g_s \phi,$$
$\tau_i=1$   (protons) or -1 (neutrons),
and for electrons, $i=e$, we have
$${\cal V}_{0e}=-e {A}_{0},\,\,\,{\boldsymbol{{\cal V}_e}}=- e\,  {\mathbf
  A}, \,\,\,  M^*_e=m_e.$$

 The time evolution of the distribution functions is described by the
Vlasov equation
\begin{equation}
\frac{\partial f_{i \pm}}{\partial t} +\{f_{i \pm},h_{i \pm}\}=0, \qquad 
\; i=p,\,n,\, e,
\label{vlasov1}
\end{equation}
where $\{,\}$ denote the Poisson brackets.
 From Hamilton's equations we derive the equations describing the time
evolution of the fields $\phi$,  $V^\mu$, $A^\mu$ and the third component of
the $\rho$-field $b_3^\mu=(b_0,\mathbf{b})$, which are given in the Appendix A.

The state which minimizes the energy of asymmetric npe matter
is characterized by the distribution functions
\begin{equation}
f_{0 i \pm}= \frac{1}{1+e^{(\epsilon \mp \nu_i)/T}} \quad i=p,n
\end{equation}
with 
$ \epsilon=\sqrt{p^2+{M^*}^2}$,
%$M^*=M-g_s \phi_0,$
\begin{equation}
\nu_i=\mu_i - g_v V_0  - \frac{g_\rho}{2}\, \tau_i b_0 - e\, A_0 
\frac{1+\tau_i}{2},
\end{equation}
and
\begin{equation}
f_{0 e \pm}= \frac{1}{1+e^{(\epsilon_{0e} \mp \mu_e)/T}} 
\end{equation}
with $\epsilon_{0e}=\sqrt{p^2+m_e^2},$
and by the  constant mesonic fields which obey the following equations
$$ m_s^2\phi_0 + \frac{\kappa}{2} \phi_0^2 + \frac{\lambda}{6} \phi_0^3 =
g_s\rho_s^{(0)}$$
$$m_v^2\,V_0^{(0)}\,=\, g_v j_0^ {(0)},$$
$$m_{\rho}^2\,b_0^{(0)}\,=\, \frac{g_\rho}{2} j_{3,0}^{(0)}   \;  ,$$
$$ V^{(0)}_i\,= b_i^{(0)}\,=\,A_0^{(0)}\,=\, A_i^{(0)}\,=\, 0.$$

Collective modes in the present approach correspond to small oscillations
around the equilibrium state. These small deviations are described by the 
linearized equations of motion and, therefore, collective modes are given as 
solutions of those equations. To construct them, let us define:
$$
f_{i \pm}\,=\, f_{0 i \pm} + \delta f_{i \pm}\;,$$
$$\phi\,=\,\phi_0 + \delta\phi\;,$$
$$V_0\,=\, V_0^{(0)} + \delta V_0\;, \;\;\;
V_i\,=\,\delta V_i\;,$$
$$ b_0\,=\, b_0^{(0)} + \delta b_0\;, \;\;\;
b_i\,=\,\delta b_i\;,$$
\begin{equation}
A_0\,=\, \delta A_0\;, \;\;\;
A_i\,=\,\delta A_i.
\end{equation}
As in \cite{npp91,previous,instabilities2} we
express the fluctuations of the distribution functions in terms of the  generating 
functions
$$S_{\pm}({\mathbf r},{\mathbf p},t)=
\mbox{diag}\left(S_{p \pm},\,S_{n \pm},\,S_{e \pm}\right),$$
such that 
$$\delta f_\pm=\{S_\pm,f_{0\pm}\}=\{S_\pm,p^2\}\frac{df_{0\pm}}{dp^2}.$$
In terms of the generating functions, the linearized Vlasov equations
for $\delta f_{i\pm}$
$$\frac{d\delta f_{i\pm}}{d t}+ \{\delta f_{i\pm}, h_{0i\pm}\}
 +\{f_{0i\pm},\delta h_{i\pm} \}=0$$
 are equivalent to the following time evolution equations
\begin{eqnarray}
  \label{eq:deltafe}
 \frac{\partial S_{e \pm}}{\partial t}& +& \{S_{e \pm},h_{0e \pm}\} =
\delta h_{e \pm} = -e\left[ \delta{A}_{0} \mp\frac{{\bf p}
  \cdot \delta{\mathbf A}}{\epsilon_{0e}}\right],\nonumber
\end{eqnarray}
\begin{equation}
  \label{eq:deltaf}
 \frac{\partial S_{i \pm}}{\partial t} + \{S_{i \pm},h_{0i \pm}\} =
\delta h_{i \pm} = \mp
\frac{g_s\,M^*}{\epsilon}\delta\phi + \delta{\cal V}_{0i} \mp\frac{{\bf p}
  \cdot \delta \boldsymbol{\cal V}_i}{\epsilon},
\end{equation}
$i=p,n$, where
$$\delta{\cal V}_{0i}=g_v \delta V_0 + \tau_i \frac{g_{\rho}}{2}\, 
\delta b_0 + e\, \frac{1+\tau_{i}}{2}\,\delta A_0,$$
$$\delta \boldsymbol{\cal V}_i= g_v \delta {\mathbf V} + \tau_i
 \frac{g_{\rho}}{2} \,\delta {\mathbf b}+ e\, \frac{1+\tau_{i}}{2}\,
\delta {\mathbf A},$$
with $ h_{0 i\pm}\,=\pm \epsilon +{\cal V}^{(0)}_{0 i}$, $i=p,n$  and 
$ h_{0 e\pm}\,=\pm \epsilon_{0e}$ .
The linearized equations of the fields are shown in the Appendix B.

Of particular interest on account of their physical relevance are
the longitudinal modes, with momentum ${\bf k}$ and frequency $\omega$,
described by the ansatz
$$
\left(\begin{array}{c}
S_{j\pm}({\bf r},{\bf p},t)  \\
\delta\phi  \\ 
\delta B_0 \\ \delta B_i
\end{array}  \right) =
\left(\begin{array}{c}
{\cal S}_{\omega\pm}^j (p,{\rm cos}\theta) \\
\delta\phi_\omega \\
\delta B_\omega^0\\ \delta B_\omega^i 
\end{array} \right) {\rm e}^{i(\omega t - {\bf k}\cdot
{\bf r})} \;  ,
$$
where $j=e,\, p,\, n$, $B=V,\, b, A$ represents the vector  fields and $\theta$ is the angle between ${\bf p}$ and ${\bf k}$. For these modes,
we get $\delta V_\omega^x = \delta V_\omega^y =0\,$, $\delta b_\omega^x =
\delta b_\omega^y = 0\,$ and $\delta A_\omega^x = \delta A_\omega^y
=0\,$. 
Calling $\delta V_\omega^z = \delta V_\omega$, $\delta b_\omega^z = \delta 
b_\omega$ and $\delta A_\omega^z = \delta A_\omega$, we will  have
$$ \delta {\cal V}_{i,z}= \delta {\cal V}_\omega^i{\rm e}^{i(\omega t - {\bf k}\cdot
{\bf r})}, \qquad  \delta {\cal V}_{0i}= \delta {\cal V}_\omega^{0i}{\rm e}^{i(\omega t - {\bf k}\cdot
{\bf r})},$$
and 
 the equations of motion 
become
\begin{equation}
i(\omega \mp \frac{k p \cos \theta}{\epsilon_{0e}} ){\cal S}_{\omega \pm}^e\,=\,-e\left[ 
\delta {A}_\omega^{0} \mp\frac{p \cos \theta}{\epsilon_{0e}} \delta{A}_\omega \right] \; ,
\label{eqs1e}
\end{equation}
\begin{equation}
i\left (\omega \mp \frac{k p \cos \theta}{\epsilon}\right )
{\cal S}_{\omega \pm}^i\,=\, \mp g_s \frac{M^*}{\epsilon}
\delta\phi_\omega  \mp \frac{p \cos \theta}{\epsilon} \delta {\cal V}_\omega^i
+ \delta {\cal V}_{ \omega}^{0i}
\label{eqs1}
\end{equation}
\begin{equation}
\left( \omega^2-k^2-m^2_{s,eff}\right) \delta\phi_\omega \,=\,
4i g_s k M^*\sum_{i=p,n}\delta\rho_{si}\label{eqphi1}
\end{equation}
\begin{equation}
\left( \omega^2-k^2-m_v^2 \right)\delta V_\omega^0\,=\,
4i g_v k \sum_{i=p,n} \delta\rho_i
\label{eqv01}
\end{equation}

\begin{equation}
\left( \omega^2-k^2-m_\rho^2 \right)\delta b_\omega^0\,=\,
4i \frac{g_\rho}{2} k \sum_{i=p,n}  \tau_i \delta\rho_i
\label{eqb01}
\end{equation}
\begin{equation}
\left( \omega^2-k^2 \right)\delta A_\omega^0\,=\,
4i e k \sum_{i=p,e} (-1)^{n_i}\delta\rho_i,
\label{eqA01}
\end{equation}
where, $ n_p=0,\,n_e=1$, and, from the continuity equation for the density currents, we get for
 the components of the vector fields
\begin{equation}
\omega\delta V_\omega^0\,=\,k\,\delta V_\omega
\label{uteis1}  
\end{equation}
\begin{equation}
\omega\delta b_\omega^0\,=\,k\,\delta b_\omega  \;.
\label{uteis2}  
\end{equation}
\begin{equation}
\omega\delta A^0\,=\,k\,\delta A_\omega  \;.
\label{uteis3}  
\end{equation}

In the above equations we have defined
\begin{equation}
m^2_{s,eff}=m_s^2+\kappa\phi_0+\frac{\lambda}{2}\phi_0^2
+g_s^2{d\rho_s^0},
\label{mseff}
\end{equation}
with
$${d\rho_s^0}= \frac{2}{(2\pi)^3}\int d^3p\,\left[ (f_{0i+}+
f_{0i-}) \frac{p^2}{\epsilon_i^3}\right],
$$
$$\delta\rho_{si}= \int \frac{d^3p}{(2 \pi)^3} \frac{p \cos \theta}
{\epsilon} \left[S_{\omega i+} \frac{df_{0i+}}{dp^2} +
%  (+\leftrightarrow -)
S_{\omega i-} \frac{df_{0i-}}{dp^2} 
\right],
$$
$$\delta\rho_{i}=\int \frac{d^3p}{(2 \pi)^3} p \cos \theta
\left[S_{\omega i+} \frac{df_{0i+}}{dp^2} -
S_{\omega i -} \frac{df_{0i-}}{dp^2} \right],
$$

$$ \frac{df_{0i \pm}}{dp^2} = \frac{1}{2 \epsilon T} f_{0i \pm}
(f_{0i \pm} -1).$$

\section{Solutions for the eigenmodes and the dispersion relation}

The solutions of eqs. (\ref{eqs1e})---(\ref{uteis3}) form a complete set of 
eigenmodes which may be used to construct a general solution for an arbitrary
longitudinal perturbation. 

Substituting the set of equations (\ref{eqphi1}-\ref{uteis3}) into
(\ref{eqs1e}) and (\ref{eqs1}) we get a set of equations for  the
unknowns $S_{\omega \pm}^{i}$ which lead to the following matricial system
$$ \left(\begin{array}{ccccc}
a_{11}&a_{12}&a_{13}&a_{14}&a_{15}\\
a_{21}&a_{22}&a_{23}&a_{24}&0\\
a_{31}&a_{32}&a_{33}&a_{34}&a_{35}\\
a_{41}&a_{42}&a_{43}&a_{44}&0\\
0&0&a_{53}&0&a_{55}\\
\end{array}\right)
\left(\begin{array}{c}
\rho^S_{\omega p} \\
\rho^S_{\omega n} \\
\rho_{\omega p} \\
\rho_{\omega n} \\
\rho_{\omega e}
\end{array}\right)=0. \label{matriz}$$
The coefficients $a_{ij}$ are defined in the Appendix C. The
  amplitudes $\rho^S_{\omega i}$ and $\rho_{\omega i}$ are  functions
  of the quantities  $S_{\omega \pm}^{i}$ given in Appendix C,
  Eqs. (\ref{rhowi}) and (\ref{rhoswi}).
The dispersion relation is obtained from the determinant of the matrix of
the coefficients, namely
$$ \left|\begin{array}{ccccc}
a_{11}&a_{12}&a_{13}&a_{14}&a_{15}\\
a_{21}&a_{22}&a_{23}&a_{24}&0\\
a_{31}&a_{32}&a_{33}&a_{34}&a_{35}\\
a_{41}&a_{42}&a_{43}&a_{44}&0\\
0&0&a_{53}&0&a_{55}\\
\end{array}\right|=0.$$
In the next section we discuss the behavior of the
ratios of amplitudes ${\rho_{\omega n}}/{\rho_{\omega p}}$ and
${\rho_{\omega e}}/{\rho_{\omega p}}$ at or inside the spinodal region 
as a function of the density and the isospin asymmetry.
\hfill

\section{Numerical results and discussions}

\begin{figure}[thb]
\begin{center}
\includegraphics[width=6.5cm,angle=0]{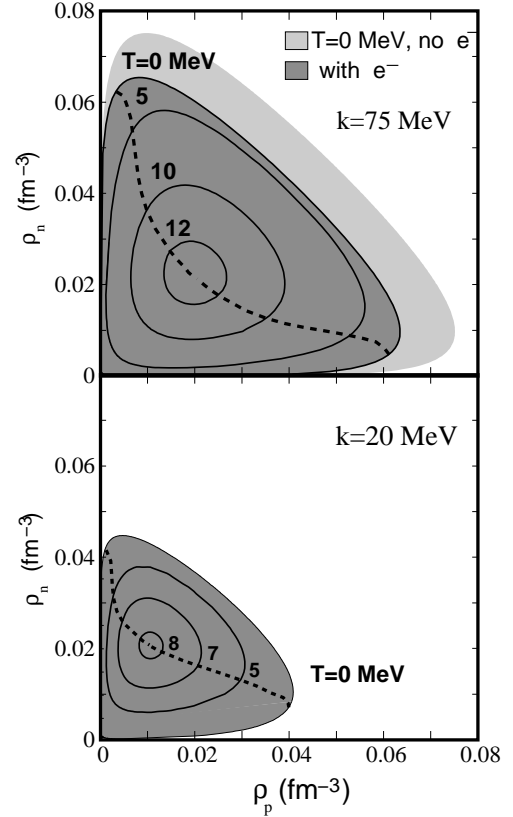}\\
\end{center}
\caption{Spinodal sections for several temperatures and momentum
  transfer $k=75$ MeV and $k=20$ MeV. }
\label{spin}
\end{figure}

In this section we present the results for the modes that  define the
instabilities of the system, corresponding to imaginary
frequencies. We are interested in understanding the effect of
the temperature and of the presence of electrons on these instabilities.

In Fig. \ref{spin} we plot the spinodal region for different temperatures and
$k=20,\, 75$ MeV. The second value of $k$ approximately defines the envelope of
the  spinodal region.  We have also considered a smaller value of $k$
in order to get a larger effect of the electrons.  As can be seen
in Fig. \ref{spint10}, the influence of the electrons
decreases, as $k$ increases up to $k\lesssim  70-80$ MeV/c (thick
lines), similar to what happened at zero temperature.
This is
expected since the Coulomb contribution varies with the inverse of the
momentum square, becoming weaker at large $k$. For higher values of $k$,
this effect becomes negligible and we recover the behavior already observed
in \cite{previous1,previous}, i.e., the instability region decreases with the increase
in the momentum transfer (thin lines). This is in fact coming from the
finite range of the nuclear interaction which reduces the binding of the
matter with a $k^{2}$ term and leads to a reduction of the spinodal region
at large $k$. 
In a thermodynamical calculation of the spinodal for npe matter
 within the NL3 parametrization done in \cite{instabilities2} it was
shown that  above 2.4 MeV the instability region completely
disappeared. In fact it is important to include electrons in a
dynamical way and allow independent electron and proton density
fluctuations in order to get the real behavior of the system.
 For
$k=75$ MeV it is seen that an instability region exists for quite high
temperatures,  $T_{cr}=12.39$ MeV. The largest critical temperature
is a function  of $k$ and, as we discuss next,  occurs for symmetric
matter only for matter without electrons or for very large values of
$k$, when the Coulomb field is negligible. The spinodals with $k=20$ MeV
  present several differences with respect to the $k=75$ MeV case: the
  critical temperature is lower, the range of densities corresponding
  to the instability region is smaller and,  while for { $k=75$ MeV
  the spinodals are practically symmetric with respect to 
  equal proton/neutron concentrations, for $k=20$ MeV the spinodal sections
  have a larger contribution in the $\rho_p<\rho_n$ range.}

\begin{figure}[t]
\begin{center}
\begin{tabular}{c}
\includegraphics[width=6.5cm,angle=0]{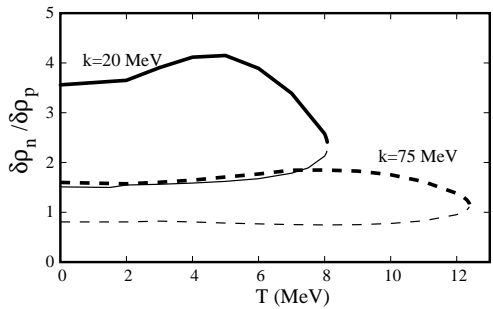}\\
\end{tabular}
\end{center}
\caption{ The ratio  $\frac{\delta\rho_n}{\delta\rho_p}$ at the
  critical points. Thin (thick) lines represent the critical points at
  the section  $\rho_p>\rho_n$ ($\rho_p<\rho_n$).
}
\label{new}
\end{figure}

\begin{figure}[b]
\begin{center}
\begin{tabular}{c}
\includegraphics[width=6.5cm,angle=0]{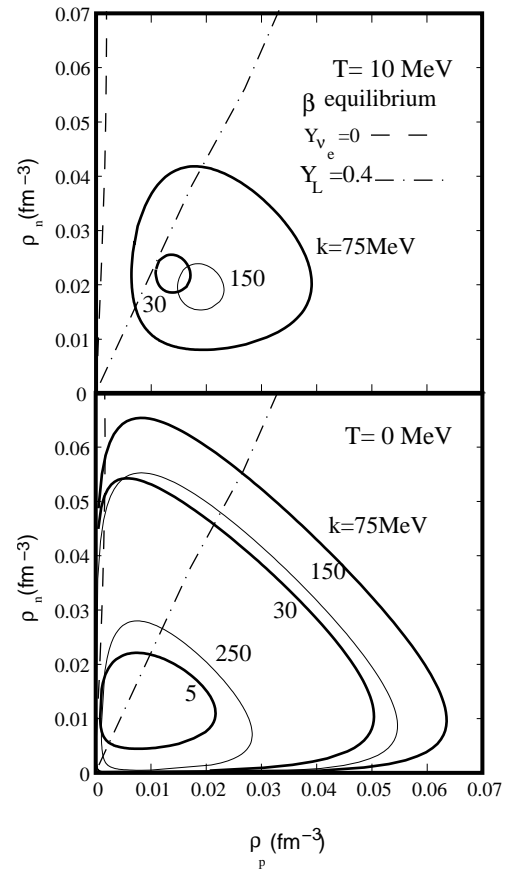}
\end{tabular}
\end{center}
\caption{Spinodal sections at T=0 and 10 MeV. The EoS for
  $\beta$-equilibrium matter is also shown.}
\label{spint10}
\end{figure}

In Fig. \ref{spin}  we also include the line of critical
points. These points of the spinodal are such that  the associated
eigenvector of the equations of motion  is tangent to the spinodal
curve. A very similar method was obtained to determine the critical 
points of the thermodynamical spinodal in \cite{instabilities2}.
Again, it is clear the  difference between the two values of $k$
  discussed. In Fig. \ref{new} we plot the ratio
  $\frac{\delta\rho_n}{\delta\rho_p}$ at which the critical point
  occurs. The lower branches (thin curves) for both cases correspond to
  the $\rho_p>\rho_n$ section while the upper branches (thick curves)
  to the  $\rho_p<\rho_n$ section. For $k=75$ MeV the ratio of
  proton/neutron density fluctuations is practically parallel to the
  line of equal concentration of protons and neutrons ($\rho_n/\rho_p=1$), 
not much different
  from  the behavior of np matter \cite{chomaz}. However
for $k=20$ MeV the eigenvectors at the instability border are tilted
in the neutron direction, preventing a strong distillation effect.

\begin{figure}[b]
\begin{center}
\begin{tabular}{c}
\includegraphics[width=6.5cm,angle=0]{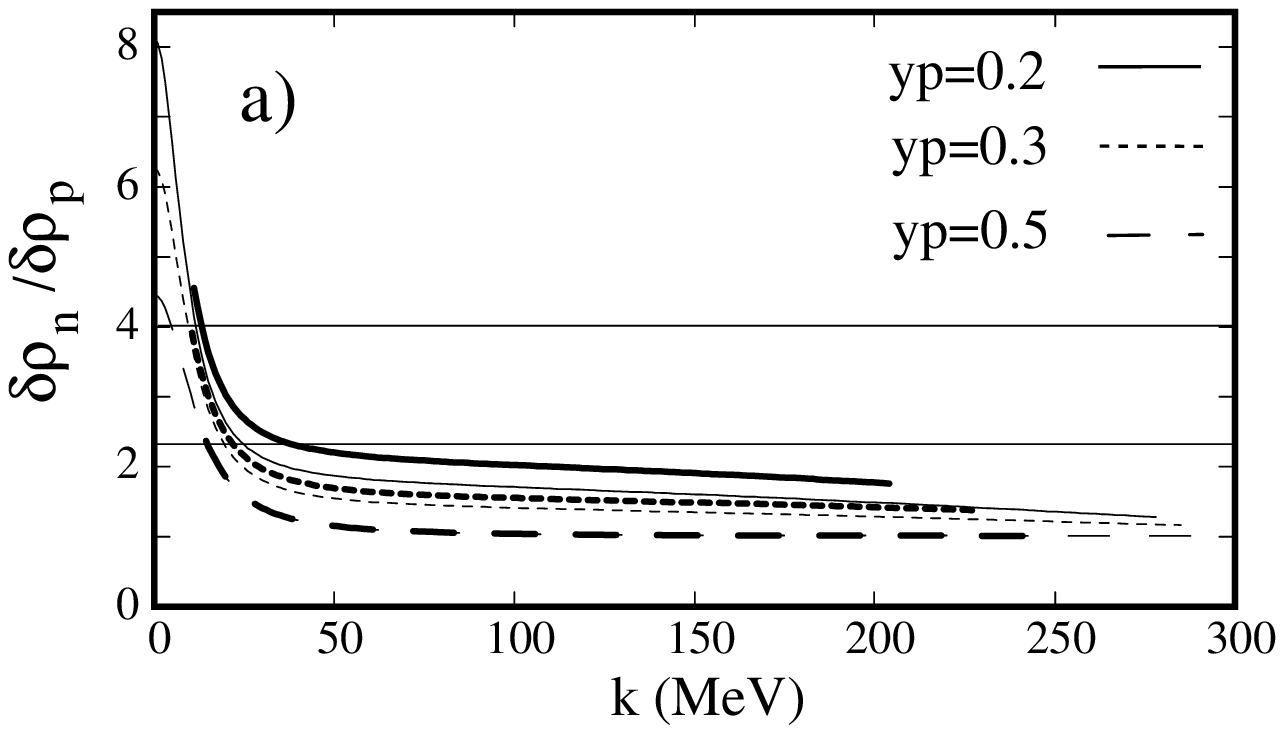}\\
\includegraphics[width=6.5cm,angle=0]{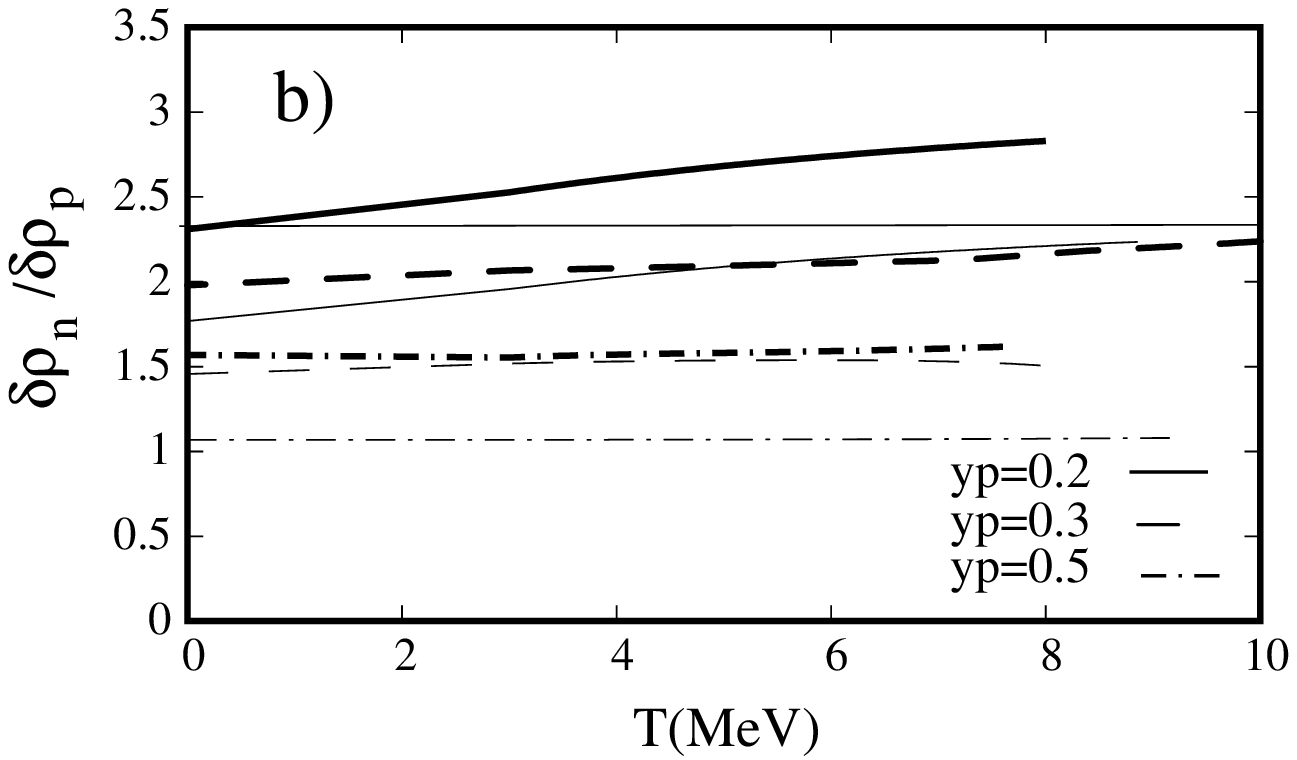}\\
\end{tabular}
\end{center}
\caption{The ratio of the neutron-density fluctuation over the
proton-density fluctuation is plotted for $\rho=0.15\rho_0$ as a
function of a)  momentum for $T=0$ (thick) and 5 MeV (thin); b) temperature for
$k=15$ (very thick line), 25 (thick line) and 75 MeV (thin line). In
this figure the dotted line represents $\rho_n/\rho_p$ for $y_p=0.3$.}
\label{fig3}
\end{figure}

As already referred, in Fig. 3 we show the spinodal regions for different 
values of $k$ at
$T=0$ and 10 MeV. We also include the equation of state (EoS) of
stellar matter in $\beta$-equilibrium without neutrinos ($Y_{\nu_e}=0$) and with
trapped neutrinos ($Y_{L}\not=0$). In this last case we have considered a 
constant leptonic fraction (electrons plus neutrinos) equal to $Y_L=0.4$, 
according to \cite{prak97}.

For T= 10 MeV, an unstable region only exists for $k \gsim \, 30$ MeV and
$k\lesssim 150$ MeV.
The spinodal region is
strongly asymmetric at small $k$ because of the electron stabilizing effect,
it becomes almost symmetric at large $k$ with the reduction of the coupling
with electrons. For T= 0 MeV it is also seen the strong asymmetry of
the spinodal at k=5 MeV as compared to k=250 MeV when the effect
of electrons is negligible.

\begin{figure}[t]
\begin{center}
\begin{tabular}{c}
\includegraphics[width=6.5cm,angle=0]{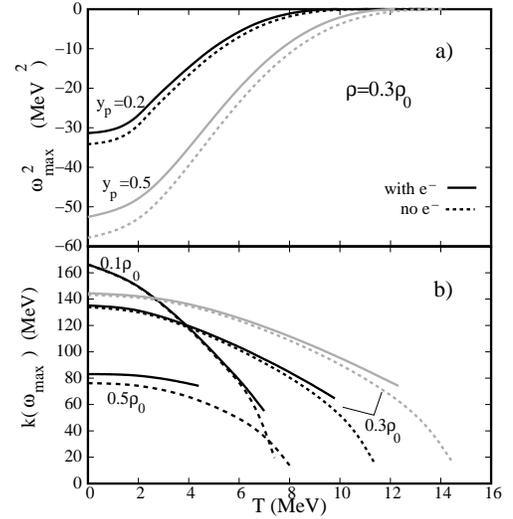}
\end{tabular}
\end{center}
\caption{Most unstable mode as a function of temperature for: 
a) the square of the corresponding frequency at $\rho=0.3 \rho_0$ and
$y_p=0.2$ (black)  and $y_p=0.5$ (grey);  b) the momentum transfer at which the most
unstable mode occurs for several densities and proton fractions (same
colors as in a) ).}
\label{wmax}
\end{figure}

The EoS of stellar matter in $\beta$ equilibrium crosses all the  spinodal
sections above $k$=6 MeV and below $k$=150 MeV at $T$=0 MeV indicating 
the existence of a nucleation phase at the crust of the star. The size of the 
inhomogeneities is of the order 8-200 fm. However, from
  Fig. \ref{wmax} we see that the most unstable mode has $k=145$ MeV
  which corresponds to a modulation with $\lambda\sim 9$ fm.
At $T$=10 MeV the $\beta$ equilibrium EoS does not cross the unstable region. 
At this temperature matter is homogeneous for all
  densities. However, for matter with trapped neutrinos the
    inhomogeneities are again present. In this case, the EoS 
crosses the unstable region even at $T$=10 MeV for  30
  MeV$<k<137$ MeV and at $T$=0 MeV the
  spinodals are crossed for all values of $k$ for which the instabilities 
occur.

\begin{figure}[b]
\begin{center}
\begin{tabular}{c}
\includegraphics[width=6.5cm,angle=0]{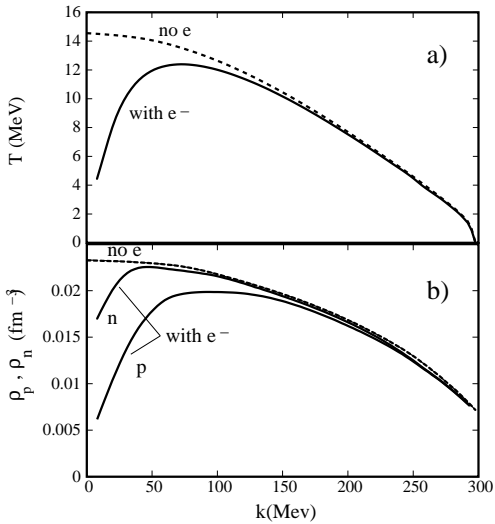}
\end{tabular}
\end{center}
\caption{Critical  a) temperatures, b) proton and
  neutron densities,  as a function of momentum transfer including/not 
including electrons.}
\label{crit}
\end{figure}

In Fig. \ref{fig3}a) the ratio of the neutron-density fluctuation over the
proton-density fluctuation is plotted as a function of momentum for
 $T=0$ and 5 MeV and $y_p=0.5,\, 0.3,\, 0.2$ and $\rho=0.15\rho_0$.  Comparing 
the finite
 temperature results with the zero temperature ones we conclude: a) for
 large $k$ the fluctuations are closer to the ratio $\rho_n/\rho_p$
 and therefore, the distillation effect is not so efficient; b) the
 anti-distillation effects,
$\frac{\delta\rho_n}{\delta\rho_p}> \frac{\rho_n}{\rho_p}$, settles in at 
larger values of $k$. This becomes more clear from Fig.
\ref{fig3}b) and it is more pronounced for asymmetric matter. In
particular for $y_p=0.3$ the ratio $\delta\rho_n/\delta\rho_p$ comes
close to the value $\rho_n/\rho_p=2.33$ (denoted by the dotted line)
for $k=25$ MeV and is completely above that value for   $k=15$ MeV. It
is also interesting to notice that for $k=15$ MeV, contrary to  $k=25,
\, 75$ MeV, asymmetric matter ($y_p=0.3$) is more unstable at higher 
temperatures than symmetric matter. This point is  discussed below together 
with Fig. \ref{crit}. 

The most unstable mode (the one with the largest
modulus of the imaginary frequency), indicating the existence of an
instability, defines the behavior of the system, namely, the most probable 
size of
the inhomogeneities which are formed due to the perturbation. In
Fig. \ref{wmax} we plot $\omega_{max}^2$ and  
the momentum $k$ at which this mode occurs as a functions of temperature and
isospin asymmetry. The modulus of $\omega_{max}^2$ decreases with temperature, 
isospin asymmetry and the presence of electrons. This behavior is expected 
since any of these factors decreases the instability region of the system.

In Fig. \ref{wmax}b) we plot $k(\omega_{max})$ for the values
of Fig. \ref{wmax}a) ($\rho=0.3\rho_0$ and $y_p=0.5, \,0.2$)  and 
for  $y_p=0.2$, $\rho=0.1\rho_0$ and $0.5\rho_0$.
For the same density, the value of $k(\omega_{max})$ also decreases both with the increase of 
temperature and isospin asymmetry.  
This means that the effect of the electrons,
which is more important for smaller $k$ values, is  more effective at higher 
temperature values and larger asymmetries. For a given temperature, electrons 
shift the most unstable mode to slightly larger values of $k$. 
If we fix the proton fraction, $y_p=0.2$, and vary the density we
conclude that close to the critical temperature the $k$ lies always in
the interval 60-80 MeV. The large difference occurs at zero
temperature: in this case the smaller the density the larger the value
of $k$.
%(This effect is stronger at larger temperatures.)
These results  seem to indicate that only under special conditions the
electrons have a strong effect 
on the behavior of the system, since the most unstable mode has 
$k> 60$ MeV and larger electron effects are expected for 
$k$ below this value as discussed next.  

In Fig. \ref{crit} we plot the critical temperatures and densities as a 
function of the wave number of the perturbation for npe matter and
nuclear (np) matter. For values of $k>150$ MeV the
critical temperatures almost coincide. As expected,  electrons
decrease drastically the critical temperatures for small values of
$k$. Another effect is a reduction of the critical density at low $k$
transfer which is more pronounced on the proton density than on the
neutron density. A high proton density region will have a larger
contribution from electrons which stabilize matter.
We conclude that while at np matter the critical temperature
occurs for symmetric matter, this is generally not anymore true if electrons
  are present. In fact,   for large values of $k$ ($k>150$ MeV) matter is 
almost symmetric at $T=T_{cr}$, but for small values of $k$ we can have
significant asymmetries for $yp\sim 0.25 - 0.4$ at the critical temperatures.

\section{Conclusions}

In the present work, we have looked at the dynamical 
instabilities of nuclear neutral warm matter which is of interest for the 
study of neutron stars and supernovae. In particular, we have studied the 
effect of the temperature and the inclusion of the Coulomb interaction on the 
instability region  of nuclear matter. The calculations were performed
within a relativistic mean-field approach to nuclear matter, namely the NL3
parametrization of the NLWM but we believe that the main conclusions
with respect to the dynamical instabilities would be similar if obtained 
within other models
with constant couplings taking into account the results of \cite{previous1}.

We may interpret the present calculation as a way of going beyond a 
total screening effect calculation which considers
 equal distributions of electron and proton clouds. On the other hand,
 no electron screening would correspond to  an homogeneous distribution
 of electrons.  We have seen
that for small $k$ transfers, electrons and protons stick together
which corresponds to a large screening effect. On the other hand for large
momentum transfers protons and electrons behave independently and this
corresponds to the no screening limit.

For np matter it was shown that {the largest} critical
   temperatures above which nuclear matter is stable occurs for
   symmetric matter. This is not anymore true for npe matter: critical
   temperatures occur for asymmetric matter with $yp<0.5$. The
   critical proton fraction can be as low as 0.22 for $k=8$ MeV, where
   $T_{cr} \simeq 4.5$ MeV. 

In \cite{wata04} the structure of hot dense matter at subnuclear
     densities has been investigated by quantum molecular dynamics (QMD) 
simulations. In this
     work the critical temperature for the phase separation is $\gsim 6$ MeV 
for the proton fraction $y_p =
0.5$ and $\gsim 5$ MeV for $y_p = 0.3$. These results seem to be compatible 
with Fig \ref{crit}. In our approach we get for $y_p=0.3$ a critical
temperature of 5 MeV at $k=10$ MeV. The maximum critical temperature is 12.38 MeV and occurs
for $y_p=0.47$ at $k=75$ MeV.  For npe matter with $y_p>0.495$ the critical temperature occurs
at $T\lesssim 8.5$ MeV for $k> 180$ MeV.   

{ We have seen that stellar matter with trapped neutrinos  is more 
affected by the
  presence of a non-homogeneous phase than neutrino free stellar  matter.}
  We have also concluded that at finite temperature 
 the distillation effect in asymmetric matter  is not so efficient, and in 
particular the anti-distillation effect defined as 
$\frac{\delta\rho_n}{\delta\rho_p}> \frac{\rho_n}{\rho_p}$, which
occurs at small $k$,  settles in at 
larger values of $k$. We may expect this effect to have consequences in warm 
stellar matter. In a supernovae explosion 99\%
of the energy of the collapse is radiated away in neutrinos. Neutrinos 
interact strongly with
neutrons because of the large weak vector charge of the neutron. {
Therefore, the way neutrons clusterize is important to
determine the neutrino mean free path}. These
 neutrinos may couple strongly to the neutron-rich matter 
low-energy modes  present in this explosive environment \cite{ksg}, and in 
this way revive the stalled supernovae shock.

The wavelength of the modulations which
   correspond to the most unstable modes increases with
   temperature. While for $T=0$ MeV the presence of electrons has
   almost no effect due to the high values of $k$ ($\sim 125-145$ MeV),
for asymmetric matter close to the critical temperatures they really 
make some difference. These are temperatures and proton fractions
of the order of the ones found in  stellar matter of protoneutron 
stars and therefore will  affect the crust formation and the neutrino 
diffusion during the cooling process. 

The results we have obtained are intimately related with the studies
  developed in \cite{hpbp,horo2,maru}, where the nuclear pasta is generally 
modeled as a charge neutral system of protons, neutrons and electrons at
  subnuclear densities. In \cite{horo2} one sees that, for small momentum 
transfers, the dynamical response
  function shows a peak, identified as a
  plasmon mode, and  displays an
  important amount of strength at low energies identified as a
  coherent density wave. In this work a fixed screening length of 10 fm
  was considered. Small momentum transfers are precisely the
  ones for which  protons are more influenced by the presence of
  electrons and therefore a carefull calculation taking into account
  this point should be carried out.

In \cite{maru} the authors have compared different treatments of
Coulomb interaction: in particular they have shown that the  full and the no
electron screening  calculations give similar results demonstrating
the weakness of the electron-screening effects. We came also to a
similar conclusion, however we could identify under which conditions
the  electron-screening effects are important, namely for small $k$ transfers.
 
In \cite{previous1} it was shown that 
density dependent hadronic models have different behaviors from the
NLWM we have studied in the present work, coming closer to
non-relativisitc  nuclear models with Skyrme forces. 
The dynamical instabilities of npe matter within these models is 
currently under investigation. The calculation of collective
  modes of neutron-rich nuclear clusters immersed in a gas of neutrons, 
including electrons dynamically,  should also be studied.

\appendix
\section{Equations for the fields}

\begin{equation}
\frac{\partial^2\phi}{\partial t^2} - \nabla^2\phi +m_s^2\phi +
\frac{\kappa}{2} \phi^2 + \frac{\lambda}{6} \phi^3
= g_s\rho_s({\bf r},t) \; ,
\label{eqmphi}
\end{equation}
\begin{equation}
\frac{\partial^2 V_0}{\partial t^2} - \nabla^2 V_0 + m_v^2 V_0\, =\,
g_v j_0({\bf r},t) \;,
\label{eqmv0}
\end{equation}
\begin{equation}
\frac{\partial^2 V_i}{\partial t^2} - \nabla^2 V_i + m_v^2 V_i\, =\,
g_v j_i({\bf r},t) \;,
\label{eqmv}
\end{equation}
\begin{equation}
\frac{\partial^2 b_0}{\partial t^2} - \nabla^2 b_0 + m_\rho^2 b_0\, =\,
\frac{g_\rho}{2} j_{3,0}({\bf r},t)\;,
\label{eqmb0}
\end{equation}
\begin{equation}
\frac{\partial^2 b_i}{\partial t^2} - \nabla^2 b_i + m_v^2 b_i\, =\,
\frac{g_\rho}{2} j_{3,i}({\bf r},t) \;,
\label{eqmb}
\end{equation}

\begin{equation}
\frac{\partial^2 A_0}{\partial t^2} - \nabla^2 A_0 \, =\,
e [j_{0p}({\bf r},t) -j_{0e}({\bf r},t)] \;,
\label{eqma0}
\end{equation}
\begin{equation}
\frac{\partial^2 A_i}{\partial t^2} - \nabla^2 A_i \, =\,
e[j_{pi}({\bf r},t) -j_{ei}({\bf r},t) ]\;,
\label{eqma}
\end{equation}

where the scalar density is
\begin{equation}
\rho_s({\bf r},t)=2\sum_{i=p,n}\int\frac{d^3p}{(2\pi)^3}\,
(f_{i+}({\bf r},{\bf p},t)+f_{i-}({\bf r},{\bf p},t))\,
\frac{M^*}{\epsilon_i}\; .
\end{equation}

The components of the baryonic four-current density are
\begin{equation}
j_0({\bf r},t)=2 \sum_{i=p,n}\int\frac{d^3p}{(2\pi)^3}\,
(f_{i+}({\bf r},{\bf p},t)-f_{i-}({\bf r},{\bf p},t))\; ,
\end{equation}
\begin{equation}
{\bf j}({\bf r},t)=2\sum_{i=p,n}\int\frac{d^3p}{(2\pi)^3}\,
(f_i({\bf r},{\bf p},t)+f_{i-}({\bf r},{\bf p},t))
\frac{{\bf p}-\boldsymbol{\cal V}_i}{\epsilon_i} \;,
\end{equation}
\begin{equation}
j_{0e}({\bf r},t)=2 \int\frac{d^3p}{(2\pi)^3}\,
(f_{e+}({\bf r},{\bf p},t)-f_{e-}({\bf r},{\bf p},t))\; ,
\end{equation}
\begin{equation}
{\bf j}_e({\bf r},t)=2\int\frac{d^3p}{(2\pi)^3}\,
(f_{e+}({\bf r},{\bf p},t)+f_{e-}({\bf r},{\bf p},t))\,
\frac{{\bf p}+e\boldsymbol{\mathbf A}}{\epsilon_e} \; ,
\end{equation}
and the components of the isovector four-current density are
\begin{equation}
j_{3,0}({\bf r},t)=2\sum_{i=p,n}\int\frac{d^3p}{(2\pi)^3}\, \tau_i
(f_{i+}({\bf r},{\bf p},t)-f_{i-}({\bf r},{\bf p},t)) \; ,
\end{equation}
\begin{equation}
{\bf j}_3({\bf r},t)=2\sum_{i=p,n}\int\frac{d^3p}{(2\pi)^3}\,
\frac{{\bf p}-\boldsymbol{\cal V}_i}{\epsilon_i}\, \tau_i \; 
(f_{i+}({\bf r},{\bf p},t)+f_{i-}({\bf r},{\bf p},t)) \; ,
\end{equation}
 with
$
\epsilon_i=\sqrt{({\bf p}-\boldsymbol{\cal V}_i)^2+{M^*}^2} \;, i=p,n
$
and
$
\epsilon_e=\sqrt{({\bf p}+e{\mathbf A})^2+m_e^2} \;.
$

\section{Linearized equations for the fields}

\begin{equation}
\frac{\partial^2\delta\phi}{\partial t^2} -
\nabla^2\delta\phi + (m_s^2+\kappa\phi_0+\frac{\lambda}{2}\phi_0^2)\delta
\phi\, =\, g_s\, \delta\rho_s \; ,
\label{phi}
\end{equation}
\begin{equation}
\frac{\partial^2\delta V_0}{\partial t^2} - \nabla^2\delta V_0 + m_v^2\delta
V_0\, =\, g_v\, \delta j_0 \; ,
\label{v0}
\end{equation}
\begin{equation}
\frac{\partial^2\delta V_i}{\partial t^2} - \nabla^2\delta V_i + m_v^2\delta
V_i\, =\, g_v\, \delta j_i \; ,
\label{vi}
\end{equation}
\begin{equation}
\frac{\partial^2\delta b_0}{\partial t^2} - \nabla^2\delta b_0 + m_\rho^2\delta
b_0\, =\, \frac{g_\rho}{2}\, \delta j_{3,0} \; ,
\label{b0}
\end{equation}
\begin{equation}
\frac{\partial^2\delta b_i}{\partial t^2} - \nabla^2\delta b_i + m_\rho^2\delta
b_i\, =\, \frac{g_\rho}{2}\, \delta j_{3,i} \; ,
\label{bi}
\end{equation}
\begin{equation}
\frac{\partial^2\delta A_0}{\partial t^2} - \nabla^2\delta A_0
\, =\, e\, \left[\delta j_{0p}- \delta j_{0e} \right]\; ,
\label{A0}
\end{equation}
\begin{equation}
\frac{\partial^2\delta A_i}{\partial t^2} - \nabla^2\delta A_i \, =\, e\,
\left[ \delta j_{pi} -\delta j_{ei}\right]\; ,
\label{Ai}
\end{equation}

with
$$
\delta\rho_s = 2\sum_{i=p,n}\int \frac{d^3 p}{(2\pi)^3}\,
\frac{M^*}{\epsilon} (\delta f_{i+}+\delta f_{i-})
-\,g_s\delta\phi~ {d\,\rho_s^0} \;,
$$
$$
\delta j_0 = 2\sum_{i=p,n}\int \frac{d^3 p}{(2\pi)^3}\, 
(\delta f_{i+}-\delta f_{i-})
%\frac{2ikM}{(2 \pi)^2 T} \sum_{i=p,n} \left(A^1_{\omega i+}-A^1_{\omega i-}
%\right),
$$
%where $A^1_{\omega i+}$ and $A^1_{\omega i-}$ are defined in equations%(\ref{a0+}) and (\ref{a0-}) respectively,
\begin{eqnarray*}
\delta {\mathbf j}& =&  2\sum_{i=p,n}\int \frac{d^3 p}{(2\pi)^3}\
\frac{\mathbf p}{\epsilon} (\delta f_{i+}+\delta f_{i-})\,\\
&-&\,2\sum_{i=p,n}\int \frac{d^3
  p}{(2\pi)^3}\,(f_{0i+}+f_{0i-})
\left(\frac{\delta\boldsymbol{\cal V}_i}{\epsilon} -
\mathbf{p}\frac{{\mathbf p}\cdot\delta{\boldsymbol{\cal V}_i}}{\epsilon^3}
\right) \;,
\end{eqnarray*}
$$
\delta j_{3,0} = 2\sum_{i=p,n}\int \frac{d^3 p}{(2\pi)^3}\,\tau_i \,
(\delta f_{i+}-\delta f_{i-}) ,
$$
\begin{eqnarray*}
\delta {\mathbf j}_3 &=&  2\sum_{i=p,n}\int \frac{d^3 p}{(2\pi)^3}\,\tau_i \,
\frac{\mathbf p}{\epsilon}(\delta f_{i+}+\delta f_{i-})\,\\
&-&\,2\sum_{i=p,n}\int \frac{d^3 p}{(2\pi)^3}\,\tau_i\, (f_{0i+}+f_{0i-})
\left(\frac{\delta\boldsymbol{\cal V}}{\epsilon}
- \mathbf{p}\frac{{\mathbf p}\cdot\delta{\boldsymbol{\cal V}}}{\epsilon^3_{0}}
\right) \;.
\end{eqnarray*}
 $$
\delta j_{0e} = 2\int \frac{d^3 p}{(2\pi)^3}\,
(\delta f_{e+}-\delta f_{e-})\; ,$$
\begin{eqnarray*}
\delta {\mathbf j}_e &=&  2\int \frac{d^3 p}{(2\pi)^3}\,\,
(\delta f_{e+}+\delta f_{e-})\,\frac{\mathbf p}{\epsilon_{0e}}\,\\
&+&
2e\int \frac{d^3 p}{(2\pi)^3}\,\, (f_{0e+}+f_{0e-})
\left(\frac{\delta{\mathbf A}}{\epsilon_{0e}}
- \mathbf{p}\frac{{\mathbf p}\cdot\delta{{\mathbf A}}}{\epsilon^3_{0e}}
\right) \;.
\end{eqnarray*}

\section{Dispersion relation}
The dispersion relation is obtained from the determinant of the
following set of five equations
$$ \left(\begin{array}{ccccc}
a_{11}&a_{12}&a_{13}&a_{14}&a_{15}\\
a_{21}&a_{22}&a_{23}&a_{24}&0\\
a_{31}&a_{32}&a_{33}&a_{34}&a_{35}\\
a_{41}&a_{42}&a_{43}&a_{44}&0\\
0&0&a_{53}&0&a_{55}\\
\end{array}\right)
\left(\begin{array}{c}
\rho^S_{\omega p} \\
\rho^S_{\omega n} \\
\rho_{\omega p} \\
\rho_{\omega n} \\
\rho_{\omega e}
\end{array}\right)=0 \label{matriz2}.$$
The coeficients $a_{ij}$ are given by
$$a_{11}=1
+c_s\left(  I_{\omega  +}^{0p}- I_{\omega  -}^{0p} \right),$$
$$a_{21}=c_s\left(  I_{\omega  +}^{0p}- I_{\omega  -}^{0p} \right),\,$$
$$a_{31}=-(c_v+c_\rho+c_e) \left(I_{\omega  +}^{1p}+I_{\omega
    -}^{1p}\right),\,$$
$$a_{41}=-(c_v-c_\rho) \left(I_{\omega  +}^{1p}+I_{\omega -}^{1p}\right),$$
$$a_{51}= c_e \left(I_{\omega  +}^{1p}+I_{\omega -}^{1p}\right)$$

$$a_{21}=c_s\left(  I_{\omega  +}^{0n}- I_{\omega  -}^{0n} \right),\,$$
$$a_{22}=1+c_s\left(  I_{\omega  +}^{0n}- I_{\omega  -}^{0n} \right),\,$$
$$a_{23}=-(c_v-c_\rho) \left(I_{\omega  +}^{1n}+I_{\omega
    -}^{1n}\right),\,$$
$$a_{24}=-(c_v+c_\rho) \left(I_{\omega  +}^{1n}+I_{\omega -}^{1n}\right),\,$$
$$a_{25}=0$$

$$a_{31}=+c_s\left(  I_{\omega  +}^{1p}+ I_{\omega  -}^{1p} \right),\,$$
$$a_{32}=+c_s\left(  I_{\omega  +}^{1p}+ I_{\omega  -}^{1p} \right),\,$$
$$a_{33}=1-(c_v+c_\rho+c_e) \left(I_{\omega  +}^{2p}-I_{\omega-}^{2p}\right),\,$$
$$a_{34}=-(c_v-c_\rho) \left(I_{\omega  +}^{2p}-I_{\omega -}^{2p}\right)$$
$$a_{35}=c_e \left(I_{\omega  +}^{2p}-I_{\omega -}^{2p}\right)$$

$$a_{41}=c_s\left(  I_{\omega  +}^{1n}+ I_{\omega  -}^{1n} \right),\,$$
$$a_{42}=c_s\left(  I_{\omega  +}^{1n}+ I_{\omega  -}^{1n} \right),\,$$
$$a_{43}=-(c_v-c_\rho) \left(I_{\omega  +}^{2n}-I_{\omega-}^{2n}\right),\,$$
$$a_{44}=1-(c_v+c_\rho) \left(I_{\omega  +}^{2n}-I_{\omega -}^{2n}\right),$$

$$a_{51}=a_{52}=a_{54}=0,\,$$
$$a_{53}=c_e \left(I_{\omega  +}^{2e}-I_{\omega-}^{2e}\right),\,$$
$$a_{55}=1-c_e \left(I_{\omega  +}^{2e}-I_{\omega -}^{2e}\right)$$
where we have  defined the following quantities
$$c_s=\frac{2 G_1^2}{(2 \pi)^2 T} \frac{1}{\bar \omega^2 - 
\omega_s^2},\qquad \bar\omega_s^2=(k^2+m^2_{s,eff})/k^2,$$
$$c_v=\frac{2}{(2 \pi)^2 T} \frac{1}{\bar \omega^2 - \bar\omega_v^2} 
\left(\frac{g_v}{k}\right)^2 (1- \bar \omega^2),
\qquad \omega_v^2=(k^2 + m_v^2)/k^2,$$  
$$c_\rho= \frac{2}{(2 \pi)^2 T} \frac{1}{\bar \omega^2 - \bar\omega_\rho^2} \left(\frac{g_\rho}{2k}\right)^2 (1- \bar \omega^2), \qquad
\omega_{\rho}^2=(k^2+m_\rho^2)/k^2,$$
$$c_e= \frac{-2}{(2 \pi)^2 T} \left(\frac{e}{k}\right)^2$$
$$\bar \omega=\frac{\omega}{k}, \quad x= \frac{p \cos \theta}{\epsilon},
\quad G_1=\frac{g_s M^*}{k}.$$

$$I_{\omega_\mp}(\epsilon)= \int_{-p/\epsilon}^{p/\epsilon}~dx 
\frac{x}{\bar \omega \pm x} = \pm \left[2 \frac{p}{\epsilon} + \bar \omega ~ln
\left| \frac{\bar \omega - p/\epsilon}{\bar \omega + p/\epsilon} \right| 
\right]$$

$$I^{ni}_{\omega_\mp}= \int_{M^*}^{\infty} \epsilon^n I_{\omega_\mp}(\epsilon) 
f_{0 i \mp}( f_{0 i \mp} -1) d\epsilon.$$

The amplitudes $\rho^S_{\omega i}$ and $\rho_{\omega i}$ are given by
\beq
\rho_{\omega i}^S= A^1_{\omega i,+}- A^1_{\omega i,-},
\label{rhowi}
\eeq
\beq
\rho_{\omega i}= A^0_{\omega i,+}+ A^0_{\omega i,-}
\label{rhoswi}
\eeq
where
\beq
A_{\omega i,\pm}^n=\int_{M^*}^\infty \epsilon^n d\epsilon
\int_{-p/\epsilon}^{p/\epsilon}
dx\, x\, S_{\omega \pm}^{i}(x,p)\, f_{0i\pm}\left( f_{0i\pm}-1
\right).
\label{awi}
\eeq

 At low densities, corresponding to a negative value of the 
compressibility, the system presents unstable modes characterized by
an imaginary frequency. In order to obtain these modes, one has to 
replace $\bar \omega$ by $i \beta$. In this case,
the integrals $I_{\omega_\mp}(\epsilon)$ become
\begin{equation}
I_{\beta_\mp}(\epsilon) = \pm\left[2 \frac{p}{\epsilon}
 - 2 \beta \tan^{-1} (p/\epsilon\beta).\right]
\end{equation}

\section*{ACKNOWLEDGMENTS}

This work was partially supported by CAPES(Brazil)/GRICES (Portugal) under 
projects 100/03, BEX1038/05-2 and FEDER/FCT (Portugal) under the project  
POCI/FP/FNU/63419/2005. A. M. S. Santos would like to thank the hospitality and
friendly atmosphere provided by the Centro de F\ii sica Te\'orica of the University
of Coimbra, during his stay in Portugal.


\begin{thebibliography}{99}

\bibitem{instabilities1}S.S. Avancini, L. Brito, D.P. Menezes and C. 
Provid\^encia, Phys. Rev. C {\bf 70}, 015203 (2004).

\bibitem{previous} S.S. Avancini, L. Brito, D.P. Menezes and C. Provid\^encia,
Phys. Rev. {\bf C 71},  044323 (2005).

\bibitem{gotas} C. Provid\^encia, D.P. Menezes and L. Brito, Nucl. Phys.
{\bf A 703}, 188 (2002);
D.P. Menezes and C. Provid\^encia, Phys. Rev. {\bf C 64}, 044306 (2001); 
D.P. Menezes and C. Provid\^encia, Nucl. Phys. {\bf A 650}, 283 (1999);
D.P. Menezes and C. Provid\^encia, Phys. Rev. {\bf C 60}, 024313 (1999).

\bibitem{pasta}  D. G. Ravenhall, C. J. Pethick, and J. R. Wilson,
  Phys. Rev. Lett. {\bf 50}, 2066 (1983); M. Hashimoto, H. Seki, and
  M. Yamada, Prog. Theor. Phys.{\bf 71}, 320 (1984).
\bibitem{horo2} C. J. Horowitz, M. A. Perez-Garcia, and J. Piekarewicz, 
Phys. Rev. {\bf C 69}, 045804 (2004).
\bibitem{wata04} G. Watanabe, K. Sato, K. Yasuoka, and T. Ebisuzaki,
  Phys. Rev. C 69, 055805 (2004); 
G. Watanabe, T. Maruyama, K. Sato, K. Yasuoka, and T. Ebisuzaki, Phys.
Rev. Lett. 94, 031101 (2005). 

\bibitem{maru}T. Maruyama, T. Tatsumi, D. N. Voskresensky, T. Tanigawa, and S. Chiba, Phys. Rev. C 72, 015802 (2005).

\bibitem{hpbp} C. J. Horowitz, M. A. Perez-Garcia, D. K. Berry, and J. 
Piekarewicz, Phys. Rev. {\bf C 72,} 035801 (2005);
C. J. Horowitz, M. A. Perez-Garcia, J. Carriere, D. K. Berry, and J. 
Piekarewicz, Phys. Rev. {\bf C 70}, 065806 (2004).
\bibitem{sawyer} R. F. Sawyer, Phys. Rev. D {\bf 11}, 2740 (1975); N. Iwamoto 
and C. J. Pethick, Phys. Rev. D {\bf 25}, 313 (1982).

\bibitem{rbp} S. Reddy, G. Bertsch and M. P. Prakash, Phys. Lett. {\bf B 475}, 
1 (2000).



\bibitem{previous1} C. Provid\^encia, L. Brito, S.S. Avancini, D. P.  Menezes, 
Ph. Chomaz, Phys. Rev. {\bf C 73}, 025805 (2006).

\bibitem{ksg}  E. Khan, N. Sandulescu, and Nguyen Van Giai, Phys. Rev. 
{\bf C 71}, 042801(R) (2005).

\bibitem{npp91} M. Nielsen, C. Provid\^encia  and J. da Provid\^encia,
  Phys. Rev. C {\bf 44}, 209 (1991)

\bibitem{npp93} M. Nielsen, C. da Provid\^encia,
J. da Provid\^encia and Wang-Ru Lin, Mod.   Phys. Lett. A {\bf 10}, 919 (1994);
M. Nielsen, C. Provid\^encia  and J. da Provid\^encia, Phys. Rev. C {\bf 47}, 
200 (1993).

\bibitem{nl3} G. A. Lalazissis, J. K\"onig and P. Ring,
Phys. Rev. {\bf C 55}, 540 (1997).

\bibitem{instabilities2}  S.S. Avancini, L. Brito, Ph. Chomaz, D.P. 
Menezes, and  C. Provid\^encia, submitted to publication.

\bibitem{chomaz} J. Margueron and P. Chomaz, Phys. Rev. C {\bf 67}, 041602(R) 
(2003); P. Chomaz, M. Colonna and J. Randrup, Phys. Rep. {\bf 389},
263 (2004). 

\bibitem{prak97} M. Prakash, I. Bombaci, M. Prakash, P. J. Ellis, J. M.
Lattimer and R. Knorren, Phys. Rep. {\bf 280}, 1 (1997).
\end{thebibliography}
\end{document}